\documentclass[11pt]{article}
\usepackage{epsf,amssymb}
\newcommand{\ft}[2]{{\textstyle\frac{#1}{#2}}}

\def\rmi{{\rm i}}
\def\rmd{{\rm d}}
\def\unity{{\mathchoice {\mathrm{1\mskip-4mu l}} {\mathrm{ 1\mskip-4mu l}}
{\mathrm{ 1\mskip-4.5mu l}} {\mathrm{ 1\mskip-5mu l}}}}
\newcommand{\Zbar}{\mathbb{Z}}

\newcommand{\Red}[1]{#1}
\newcommand{\Blue}[1]{#1}
\newcommand{\ForestGreen}[1]{#1}
\newcommand\doingARLO[2][]{%
  \ifx\mmref\undefined #1\else #2\fi
}

\csname @addtoreset\endcsname{equation}{section}

\begin{document}

\begin{titlepage}
\begin{flushright}
KUL-TF-01/13\\
hep-th/0105158
\end{flushright}
\vspace{.5cm}
\begin{center}
\baselineskip=16pt {\LARGE    The scalars of $N=2$, $D=5$ \\ \vskip 0.2cm
and attractor equations.
}\\
\vfill
{\large Antoine Van Proeyen, 
  } \\
\vfill
{\small  Institute for theoretical physics K.U. Leuven\\
Celestijnenlaan 200D, B-3001 Leuven, Belgium }
\end{center}
\vfill
\begin{center}
{\bf Abstract}
\end{center}
{\small Theories in 5 dimensions with minimal supersymmetry are studied
for domain-wall solutions and in the context of the AdS/CFT
correspondence. The scalar manifold is a product of a very special real
manifold and a quaternionic-K\"{a}hler manifold. Superconformal methods can
clarify the structure of these manifolds, which are part of the family of
special manifolds. BPS solutions depending on the scalars and a warp
factor of the 5-dimensional metric with a flat 4-dimensional metric can
interpolate between critical points determined by algebraic attractor
equations. The mixing of vector and hypermultiplets is essential to
obtain UV and IR critical points.} \vspace{2mm} \vfill \hrule width 3.cm
{\footnotesize \noindent $^\dagger$ To be published in the proceedings of
the XXXVII Karpacz Winter School,  Proceedings Series of American
Mathematical Society, eds. A. Jadczyk, J. Lukierski, J. Rembielinski }
\end{titlepage}
\newpage
\section{Introduction}

Supersymmetric theories in 5 dimensions have got new interest in the
context of the AdS/CFT correspondence and for a supersymmetrisation of
the Randall--Sundrum (RS) scenario. In both cases one uses at the end a
metric of the form
\begin{equation}
\rmd s^2 = a(x^5)^2 \rmd x^{\underline{\mu}} \rmd x^{\underline{\nu}}
\eta_{\underline{\mu\nu}}  + (\rmd x^5)^2\,, \label{vpro-metric}
\end{equation}
where $\underline{\mu },\underline{\nu }=0,1,2,3$. We thus have a flat
4-dimensional space with a warp factor $a$ that depends on the fifth
direction $x^5$. We first want to draw the attention on two different
concepts, which are called either `smooth solutions' or `singular
sources'. With 'smooth solutions' we~\cite{vpro-Ceresole:2001wi} mean
that we consider the generic 5-dimensional supergravity
theory~\cite{vpro-AnnaGianguido}, and we look for a solution where the
warp factor has the required form as explained above. On the other hand,
`singular sources' means that 3-brane sources are inserted at specific
places. The fifth dimension is in that case an orbifold $S^1/\Zbar_2$,
and the sources sit at its fixed points. The 5-dimensional bulk action is
supplemented by a brane term that involves delta functions $\delta
(x^5-x^5_{\rm fixed})$ times a four-dimensional action.
We~\cite{vpro-susyd5sing} provided a simple general mechanism how to
implement supersymmetry in such a scenario despite the singularities in
spacetime due to the delta functions. It introduces a new 4-form field in
the bulk supergravity, that appears also as 4-form for the brane
Wess--Zumino term. This construction has been reviewed also
in~\cite{vpro-Bergshoeff:2000ii} and we will omit therefore this part of
the talk from these proceedings. The mechanism inspired a generalization
to 10 dimensions, leading to new formulations of type IIA (and also type
IIB) supergravity in~\cite{vpro-Bergshoeff:2001pv}, as reviewed in the
talk of Eric Bergshoeff in this school. This setup is a similar to the
Ho\v{r}ava--Witten~\cite{vpro-Horava:1996ma} theory, which was reduced to
5 dimensions in~\cite{vpro-Lukas:1998tt}. It has been considered by
various
groups~\cite{vpro-Gherghetta:2000qt,vpro-ABN00,vpro-Falkowski:2000er,vpro-Zucker:1999ej},
and our bulk\& brane solution was used for cosmology
in~\cite{vpro-Brax:2000xk,vpro-Brax:2001fh}.

We restrict ourselves here to `smooth solutions'. The main part of this
review will be devoted to the structure of the manifolds of scalars that
appear in these theories. We explain how the superconformal methods
clarify the structure, referring to new results
of~\cite{vpro-Bergshoeff:2001hc,vpro-Fujita:2001kv}, and pay special
attention to definitions of quaternionic(-K\"{a}hler) manifolds. At the end,
we consider the scalar-dependent solutions in the warped background
(\ref{vpro-metric}), and show how the critical points are determined by
algebraic attractor equations~\cite{vpro-Ceresole:2001wi}, generalizing
earlier similar equations with vector multiplets in 4
dimensions~\cite{vpro-Ferrara:1995ih,vpro-Strominger:1996kf,vpro-Ferrara:1996dd}
to include hypermultiplets. To analyse the properties of the critical
points, a general formula on the scalar mass matrix~\cite{vpro-ACCVP}
gives a lot of insight.

\section{$N=2$, $D=5$ supergravity and its scalars}

Pure $N=2$, $D=5$ supergravity\footnote{$N=2$ means 8 real supercharges:
4-component spinors in an $SU(2)$ doublet. It is minimal supersymmetry in
5-dimensional Minkowski space. Note that for other signatures (2 or 3
time directions) one can impose Majorana conditions such that only 4 of
them survive ($N=1$).} contains a graviton, two gravitini and a
graviphoton (spin 1). The theory can contain vector multiplets, each
containing a vector, a doublet of spinors and a scalar. These scalars
define a `very special real manifold', as we will explain. Furthermore,
there are hypermultiplets each containing 2 spinors and 4 scalars. The
latter define a quaternionic-K\"{a}hler manifold. Note that we will not
consider tensor multiplets. The antisymmetric tensors are dual to vectors
in 5 dimensions. That duality is only for Abelian couplings, but that is
all that we consider here. Remark, however, that for non-Abelian theories,
antisymmetric tensor multiplets lead to more general possibilities, as
discussed in detail in~\cite{vpro-GunZag}.

We first review the superconformal construction of matter couplings in
supergravity. We then discuss the two scalar manifolds. Finally we discuss
the consequences of gauging isometries of the quaternionic manifold using
the vector fields.

\subsection{Superconformal tensor calculus}

Superconformal tensor calculus provides a way to construct matter-coupled
super-Poincar\'{e} theories. The aim here is thus not to end up with a
superconformal theory, but rather to provide a way to construct the
general supergravity theories. Using conformal invariance facilitates the
construction of the theory and leads to more insight in the structure of
the theory. E.g. the structure of special K\"{a}hler manifolds was developed
by using superconformal tensor calculus.

The method thus consists of first constructing a theory invariant under
the superconformal group, and then fixing all the invariances that are
not required for a super-Poincar\'{e} invariant theory. A superconformal
group consists of the conformal group (with translations $P_\mu $,
Lorentz rotations $M_{\mu \nu }$, dilatations $D$, and special conformal
generators $K_\mu $), supersymmetry $Q$, and a special supersymmetry $S$,
and finally some $R$-symmetry. The latter is a bosonic group that acts on
the supersymmetries and appears in the anticommutator of ordinary and
special supersymmetry. In our case, this group is $SU(2)$, and the full
superconformal algebra defines the supergroup $F^2(4)$ (where the index
'2' indicates the particular real form of the superalgebra, see table~5
in~\cite{vpro-VanProeyen:1999ni}). Its bosonic subgroup is $SO(5,2)\times
SU(2)$, where the first factor is the conformal group and the second is
the $R$-symmetry group.

Note that in this case the bosonic subgroup is a direct product of the
conformal group and the $R$-symmetry group. This is the case in the
superconformal algebras classified by Nahm~\cite{vpro-Nahm:1978tg}. It
implies that bosonic symmetries that are not in the conformal algebra are
spacetime scalars. This is not a necessity. Other examples have been
considered first in 10 and 11 dimensions in~\cite{vpro-vanHolten:1982mx}.
Recently, a new classification has appeared in~\cite{vpro-D'Auria:2000ec}
of which we can extract table~\ref{vpro-tbl:sconfalg}
\begin{table}
\begin{tabular}{|c|lll|}
\hline
 D & supergroup & \multicolumn{2}{c|}{bosonic group}  \\
\hline
 3 &$ OSp(N|4)    $\hspace{2.7cm }\phantom{.} &$    Sp(4)=SO(3,2)
  $\hspace{2.7cm }
  \phantom{.}&$     SO(N)       $ \\
 4 &$  SU(2,2|N)  $&$    SU(2,2)=SO(4,2)   $&$   SU(N) {\times}U(1)   $ \\
 5 &$  OSp(8^*|N) $&$   SO^*(8)\supset SO(5,2)    $&$  USp(N)         $ \\
   &$    F(4)      $&$             SO(5,2)  $&$    SU(2) $ \\
 6 &$  OSp(8^*|N) $&$   SO^*(8)= SO(6,2)    $&$   USp(N)        $ \\
 7 &$  OSp(16^*|N) $&$  SO^*(16)\supset SO(7,2)    $&$ USp(N)          $ \\
 8 &$  SU(8,8|N)  $&$    SU(8|8) \supset SO(8,2) $&$    SU(N) {\times}U(1)  $ \\
 9 &$  OSp(N|32)  $&$   Sp(32) \supset SO(9,2)   $&$    SO(N)        $ \\
 10 &$OSp(N|32)    $&$ Sp(32) \supset SO(10,2)    $&$ SO(N)           $ \\
 11 &$OSp(N|64)    $&$ Sp(64) \supset SO(11,2)    $&$ SO(N)           $ \\
\hline
\end{tabular}
\caption{\it Superconformal algebras, with the two parts of the bosonic
subalgebra: one that contains the conformal algebra and  the other one is
the $R$-symmetry. In the cases $D=4$ and $D=8$, the $U(1)$ factor in the
$R$-symmetry group can be omitted for $N\neq 4$ and $N\neq 16$,
respectively.} \label{vpro-tbl:sconfalg}
\end{table}
for dimensions from 3 to 11. The bosonic subgroup contains always two
factors. One contains the conformal group. If that factor is really the
conformal group, then the algebra appears in Nahm's classification. Note
that 5 dimensions is a special case. There is a generic superconformal
algebra for any extension. But for the case $N=2$ there exists a smaller
superconformal algebra that is in Nahm's list. So far, superconformal
tensor calculus has only been based on algebras of Nahm's type. Note that
for $D=6$ or $D=10$, where one can have chiral spinors, only the case
that all supersymmetries have the same chirality has been included.

For the methods that are used in superconformal tensor calculus, we refer
to existing reviews, as the one that appeared in the Karpacz proceedings
of 1983~\cite{vpro-VanProeyen:1983wk}. A useful example is its
application in $N=1$, $D=4$
supergravity~\cite{vpro-Cremmer:1983en,vpro-Kugo:1983mr}, that has been
written down in detail in~\cite{vpro-Kallosh:2000ve}.

The basic multiplet is the one that contains the gauge field of all the
symmetries in the superconformal group. This is called the Weyl
multiplet, and has recently been constructed for $N=2$, $D=5$
in~\cite{vpro-Bergshoeff:2001hc,vpro-Fujita:2001kv}. There are two
versions, as it is the case in six
dimensions~\cite{vpro-Bergshoeff:1986mz}. Both versions have 32+32
components and are equivalent. In fact, there is a procedure to go from
one to the other~\cite{vpro-Bergshoeff:2001hc}. We will restrict
ourselves to one version, which is the one used primarily also in 4 and 6
dimensions. Its content is given in table~\ref{vpro-tbl:Weylmult}.
\begin{table}
$ \begin{array}{|l|rrr|cc|} \hline
\mbox{(gauge) field}  & D=4 & D=5 & D=6& \mbox{gauge transf.}&\mbox{subtracted}\\
\hline
\Blue{e_\mu {}^a     & 5 & 9 & 14&P^a & M_{ab},\,D\\
b_\mu    &\multicolumn{3}{c|}{\mbox{compensating } K^a} &D&K^a \\
\omega _\mu {}^{ab}& \multicolumn{3}{c|}{\mbox{composite}} &M^{ab}& \\
f_\mu {}^a & \multicolumn{3}{c|}{\mbox{composite}} &K^a&\\
V_{\mu i}{}^j   & 9 & 12 & 15& SU(2) & \\
A_\mu &3 &-&-&U(1)& \\}
\Red{\psi _\mu {}^i  & 16 & 24 & 32&Q^i& S^i \\[1mm]
\phi _\mu {}^i & \multicolumn{3}{c|}{\mbox{composite}} &S^i& }\\
\Blue{T_{ab}, T_{abc}^-  & 6 & 10 & 10& &\\
D                                  & 1 &  1 &  1&&\\} \Red{\chi ^i  & 8 &
8 & 8  &&}\\ 
 \hline
\mbox{TOTAL}  & \Blue{24}+\Red{24}\ & \Blue{32}+\Red{32}\ & \Blue{40}+\Red{40}\ && \\
\hline
\end{array}$
  \caption{\it Standard Weyl multiplet in 4, 5 and 6 dimensions.}\label{vpro-tbl:Weylmult}
\end{table}
We indicate for each field the number of components in each dimension, the
symmetry for which it is a gauge field, and possibly other gauge
transformations that have been used to reduce its number of degrees of
freedom in this counting.

Once one has this multiplet, one can add other multiplets, i.e.
representations of the superconformal algebra. In order to satisfy this
algebra, the transformation laws of the fields in these multiplets will
involve the fields of the Weyl multiplet. Then one constructs a
superconformal invariant action, and finally one has to fix the
superfluous symmetries. Remark that we already used the $K^a$ symmetry to
put the gauge field of dilatations, $b_\mu $, equal to zero. The remaining
superfluous symmetries are therefore the dilatations $D$, the special
supersymmetries $S^i$, and the $R$-symmetry $SU(2)$.

As we mentioned in the beginning of this chapter, we consider general
couplings with $n$ vector multiplets and $r$ scalar multiplets.
Table~\ref{vpro-tbl:superPoinc} gives their content, the names that we
use for the fields, and the corresponding range of indices.
\begin{table}
\begin{tabular}{|l|ccc|c|l|}
\hline
 spin & pure SG & vector mult. & hypermult. & field & indices \\
\hline
 2 & 1 &   &   & $e_\mu ^a$  & $\mu ,a=0,\ldots ,4$ \\
 $\ft32$ &   2    &  & & $\psi _\mu ^i$ &  $i=1,2$  \\
 1 & 1 & $n$ &   & $A_\mu ^I$ & $I=0,\ldots ,n$ \\
 $\ft12$ &   & $2n$ & $2r$ & $\lambda ^x_i$, $\zeta ^A$ & $A=1,\ldots ,2r$ \\
 0 &   & $n$ & $4r$ & $\phi ^x$, $q^X$ & $x=1,\ldots ,n$; $X=1,\ldots ,4r$ \\
\hline
\end{tabular}
  \caption{\it Multiplets and fields of the super-Poincar\'{e} theories}\label{vpro-tbl:superPoinc}
\end{table}
In the superconformal method, these are obtained in a different way. One
starts with the Weyl multiplet, and adds vector multiplets and
hypermultiplets in representations of the superconformal algebra. As well
for the vector multiplets as for the hypermultiplets, one starts by
adding one more multiplet than appears in the final super-Poincar\'{e}
theory. These `compensating multiplets' contain the degrees of freedom
that will be gauge-fixed. This is schematically represented in
table~\ref{vpro-tbl:fieldsConf}.
\begin{table}
\begin{tabular}{|c|ccc|ll|}
\hline
spin& Weyl  & vector  & hyper & gauge fix &auxiliary \\
\hline
2&  $e_\mu ^a$    &               &         & &\\
$\ft32$&  $\psi _\mu ^i$ &              &         &  &\\
&  $V_{\mu i}{}^j$, $T_{ab}$ &   &         & &auxiliary \\
 1&                & $n+1$      &           &  &\\
$\ft12$&$\chi ^i$ &    $2(n+1)$ & $2(r+1)$ & 2: $S$ & $\chi ^i$ with 2
others  \\
0& $D$ &            $n+1$ & $4(r+1)$ & 1: dilatations, 3: $SU(2)$& $D$ and
 1 other   \\
\hline
\end{tabular}
  \caption{\it Multiplets and fields in the superconformal construction}\label{vpro-tbl:fieldsConf}
\end{table}
It is indicated how the superfluous symmetries are fixed, and how some of
the fields of the Weyl multiplet serve as Lagrange multipliers eliminating
degrees of freedom of the spin $1/2$ and scalar fields. The field $V_{\mu
i}{}^j$ will be eliminated by its field equation, and will play the role
of $SU(2)$ curvature of the quaternionic manifold defined by the
hyperscalars. The field $T_{ab}$ will become a function of the field
strengths of the vectors in the vector multiplet (dressed by the
scalars), and plays the role of gauge field that enters in the gravitino
transformation (related to the central charge).

\subsection{Very special real and quaternionic-K\"{a}hler manifolds}
The manifolds of supergravity--matter couplings in $D=5$ are similar to
those that are known from $N=2$ in 4 dimensions.
Table~\ref{vpro-tbl:superPoinc} would be nearly identical for 4
dimensions, except that each vector multiplet then contains two scalars.
The supersymmetry defines a complex structure, and the manifold is
K\"{a}hlerian. In $N=1$ supergravity, general K\"{a}hler manifolds are possible.
In $N=2$ they are restricted to a category that is called `special K\"{a}hler
manifolds'~\cite{vpro-deWit:1984pk}. The quartets of scalars in
hypermultiplets are connected by 3 complex structures and the manifold is
quaternionic-K\"{a}hler~\cite{vpro-Bagger:1983tt}. Another recent review
containing the fundamental facts of these manifolds is given
in~\cite{vpro-Fre:2001jd}.

\subsubsection{Very special real manifolds}

We first consider the vector multiplets~\cite{vpro-Gunaydin:1984bi}. In 5
dimensions, these have real scalars (one of the scalars of 4 dimensions
sits in the $5d$-vector). We define 'very special real
manifolds'~\cite{vpro-brokensi} as those that appear in these couplings of
vector multiplets to 5-dimensional supergravity. It is clear from the
above, that they can be described in superconformal tensor calculus by
starting with $n+1$ scalars, which we denote $h^I$, as in
table~\ref{vpro-tbl:fieldsConf}. Then we impose a dilatational gauge
choice. This defines an $n$-dimensional hypersurface in the
$(n+1)$-dimensional space.

The locally supersymmetric action of the vector multiplets in 5
dimensions contains always a Chern--Simons term of the form $C_{IJK}
A^I\rmd A^J\rmd A^K$. In order for this to be gauge-invariant, the
$C_{IJK}$ have to be constant. This tensor is completely symmetric in its
indices, and supersymmetry implies that the full action is determined by
these constants (up to the choice of coordinates on the manifold). Thus
the set of numbers $C_{IJK}$ are all one needs to specify a very special
real manifold~\cite{vpro-Gunaydin:1984bi}. For an arbitrary set, one
still has to verify whether they allow a non-empty domain with
positive-definite metric on the scalar manifold.

The dilatational gauge choice that is most appropriate is the condition
\begin{equation}
  C_{IJK}h^I(\phi )h^{J}(\phi)h^{K}(\phi ) =1\,.
 \label{vpro-Chhh1}
\end{equation}
$\phi^x$ are coordinates on this manifold such that the embedding $h(\phi
)$ satisfies the condition. The metric on the scalar manifold is then
\begin{equation}
  g_{xy}= -3 (\partial _x h^I)(\partial _y h^J) C_{IJK}h^K\,.
 \label{vpro-metricvsp}
\end{equation}

\subsubsection{Quaternionic-K\"{a}hler manifolds}

Let us now look at the other side: the hypermultiplets. We first define
quaternionic manifolds. We start with a $4r$-dimensional manifold with
coordinates $q^X$. At each point there is a tangent space where the
vectors are labelled with indices $(iA)$ (see ranges in
table~\ref{vpro-tbl:superPoinc}). These are connected by $4r\times 4r$
vielbeins $f_X^{iA}$ or their inverses $f^X_{iA}$. We will here introduce
the quaternionic manifolds starting from these vielbeins. Quaternionic
manifolds entered physics in~\cite{vpro-Bagger:1983tt},
and~\cite{vpro-Galicki:1987ja} contains a lot of interesting properties.
There were two workshops on quaternionic geometry where mathematics and
physics results were brought
together~\cite{vpro-QuatWorksh1,vpro-QuatWorksh2}. Other recent papers
that review the properties of quaternionic manifolds
are~\cite{vpro-Fre:2001jd,vpro-D'Auria:2001kv}.

For supersymmetry, starting from vielbeins is a convenient approach
because these are the objects that one uses from the very beginning, i.e.
in the supersymmetry transformations of the hyperscalars:
\begin{equation}
  \delta (\epsilon ) q^X= f^X_{iA}\bar \epsilon ^i\zeta ^A\,.
 \label{vpro-delepsqX}
\end{equation}
\paragraph{Almost quaternionic manifolds}

We thus have
\begin{equation}
   f_{Y}^{iA}  f_{iA}^X=\delta _Y{}^X\,,\qquad f_{X}^{iA}  f_{jB}^X=\delta
   _j{}^i \delta _B{}^A\,.
 \label{vpro-ff1}
\end{equation}
These vielbeins satisfy a reality condition defined by matrices $E_i{}^j$
and $\rho _A{}^B$ that satisfy
\begin{equation}
  E\,E^*=-\unity _2\,, \qquad \rho \,\rho ^*=-\unity _{2r}\,.
 \label{vpro-EErhorho}
\end{equation}
One may choose a standard antisymmetric form for $\rho $ and identify $E$
with $\varepsilon $ by a choice of basis. The reality condition for the
vielbeins are
\begin{equation}
  (f_X^{iA})^*=f_X^{jB}E_j{}^i\rho _B{}^A\,.
 \label{vpro-realfrho}
\end{equation}
The transformations on variables with an $A$ index are by the reality
condition restricted to $G\ell (r,Q)=SU^*(2n)\times U(1)$.

We define complex structures as ($r =1,2,3$ and using the three sigma
matrices)
\begin{equation}
  J_X{}^{Yr }\equiv -\rmi f_X^{iA}(\sigma ^r )_i{}^jf_{jA}^Y\,, \qquad
  \Rightarrow\qquad J_X{}^Y{}_i{}^j\equiv \rmi
  J_X{}^{Yr }(\sigma ^r )_i{}^j = 2f_{X}^{jA}f_{iA}^Y-\delta _i{}^j\delta
  _X{}^Y\,.
 \label{vpro-defJf}
\end{equation}
We use the same transition between triplet and doublet notation below for
other quantities. The complex structures satisfy, due to
(\ref{vpro-ff1}), the quaternion algebra
\begin{equation}
  J^r J^s =-\unity _{4r}\delta ^{r s }+\varepsilon ^{r s t
  }J^t\,.
 \label{vpro-defJ}
\end{equation}
This defines the manifold to be `almost quaternionic'.

\paragraph{Quaternionic manifolds}

We now suppose that there is a torsionless connection $\Gamma
_{XY}^Z=\Gamma _{YX}^Z$. Consider then
\begin{equation}
\Omega _{X\,jB}{}^{iA}\equiv   f^Y_{jB}\left( \partial _X f_Y^{iA}-\Gamma
_{XY}^Zf_Z^{iA}\right)=-\omega _{Xj}{}^i \delta _B{}^A-\omega
_{XB}{}^A\delta _j{}^i\,,
 \label{vpro-Omega}
\end{equation}
where $\omega _{Xj}{}^i$ is traceless. If this $\Omega _{X\,jB}{}^{iA}$,
for each $X$, would be a general $4r\times 4r$ matrix, then we would say
that the holonomy is not restricted (or sits in $G\ell (4r)$). The
splitting as in the right-hand side of this equation implies that the
holonomy group is restricted to $SU(2)\times G\ell (r,Q)$. We can write
(\ref{vpro-Omega}) as the covariant constancy of the vielbein:
\begin{equation}
  \partial _X f_Y^{iA}-\Gamma _{XY}^Zf_Z^{iA}+f_Y^{jA}\omega _{Xj}{}^i +f_Y^{iB}\omega
_{XB}{}^A=0\,,
 \label{vpro-covconstf}
\end{equation}
with composite gauge fields for $SU(2)$ and $G\ell (r,Q)$. These
conditions promote the almost quaternionic structure to a quaternionic
structure, and the manifolds is `quaternionic'. If the $SU(2)$ connection
is zero, they are called `hypercomplex'.

The integrability condition of (\ref{vpro-covconstf}), (multiplied by a
vielbein) is
\begin{equation}
R^Z{}_{WXY}= f^Z_{iA}f_W^{jA}{\cal R} _{XYj}{}^i +f^Z_{iA}f_W^{iB}{\cal
R}_{XYB}{}^A= -J_W{}^{Zr }{\cal R} _{XY}{}^r +f^Z_{iA}f_W^{iB}{\cal
R}_{XYB}{}^A \,,
 \label{vpro-Rdecomp}
\end{equation}
where respectively the metric curvature $R^Z{}_{WXY}\equiv 2\partial _{[X}
\Gamma _{Y]W}^Z+2\Gamma ^Z_{V[X}\Gamma ^V_{Y]W} $, the $SU(2)$ curvature
${\cal R}_{XY}{}^r
\equiv  2\partial _{[X}\omega _{Y]}{}^r +2\omega _{[X }{}^s\omega
_{Y]}{}^t\varepsilon ^{rst} $, and the $G\ell (r,Q)$ curvature ${\cal
R}_{WXB}{}^A$ appear.

\paragraph{Quaternionic-K\"{a}hler manifolds}

Quaternionic-K\"{a}hler manifolds (including `hyperk\"{a}hler' for the case that
the $SU(2)$ curvature vanishes) by definition have a metric. First define
an Hermitian metric $d_{AB}$ such that $C_{AB}\equiv \rho
_A{}^Cd_{CB}=-C_{BA}$. By redefinitions of the
basis~\cite{vpro-deWit:1985px} one may diagonalize $d$ while
simultaneously bringing $C$ to a canonical form
\begin{equation}
  d=\pmatrix{\unity _{2p}&0\cr 0&-\unity _{2(r-p)}}\,,\qquad
  C=\pmatrix{0&1&&&\cr -1&0&&& \cr &&0&1&\cr &&-1&0&\cr &&&&\ddots}.
 \label{vpro-canonicalC}
\end{equation}
The subgroup of $G\ell (r,Q)$ that preserves $d$ is $USp(2p,2r-2p)$. We
define then the metric of the manifold to be\footnote{In principle one
could also introduce a metric in the $SU(2)$ part, but as there is no
choice for the signature in this sector, this is irrelevant, and we
identify $E_i{}^j=\varepsilon _{ij}$ (with $\varepsilon _{ik}\varepsilon
^{jk}=\delta _i{}^j$).}
\begin{equation}
  g_{XY}=f_X^{iA}C_{AB}\varepsilon _{ij}f_Y^{jB}\,.
 \label{vpro-defg}
\end{equation}
We use $\varepsilon _{ij}$ and $C_{AB}$ to raise and lower indices
according to the NW--SE convention
\begin{equation}
A_i=A^j\varepsilon _{ji}\,,\qquad A^i=\varepsilon ^{ij}A_j\,, \qquad
A_A=A^BC_{BA}\,,\qquad A^A=C^{AB}A_B\,.
 \label{vpro-NWSE}
\end{equation}
One can check that this raising and lowering of indices, together with
the usual raising and lowering of indices $X$ by $g_{XY}$ and its inverse
is consistent with defining $f_X^{iA}$ and $f_{iA}^X$ as each others
inverses.

The choice of the signature in (\ref{vpro-canonicalC}) is relevant for the
reality conditions, which  are then
 $ (f_X^{iA})^*d_{AB}= f_{XiB}$. Our notations hide $d$ from other places.

For $r>1$ one can prove that these manifolds are Einstein, and that the
$SU(2)$ curvatures are proportional to the complex structures:
\begin{equation}
 R_{XY}=\frac{1}{4r} g_{XY} R\,,\qquad    {\cal R}_{XY}^r =\ft12\nu  J_{XY}^r\,,\qquad
\nu  =\frac{1}{4r(r+2)}R\,.
 \label{vpro-RlambdaJ}
\end{equation}
(with $R_{XY}=R^Z{}_{XZY}$). For $r=1$ this is part of the definition of
quaternionic-K\"{a}hler manifolds. Hyperk\"{a}hler manifolds are those where the
$SU(2)$ curvature is zero, and these are thus also Ricci-flat.

\paragraph{Supergravity}

In supergravity we find all these constraints from requiring a
supersymmetric action. Moreover, we need for the invariance of the action
that the last equation of (\ref{vpro-RlambdaJ}) is satisfied with $\nu
=-1$. This implies that the scalar curvature is $R=-4r(r+2)$. This
excludes e.g. the compact symmetric spaces.

\subsubsection{The family of special manifolds}

We now place these manifolds in the context of the manifolds that are
obtained for supersymmetries with 8 real supercharges. Note that a higher
number of supercharges would restrict the possibilities for the scalar
manifolds to a discrete number of symmetric spaces. We first consider
vector multiplets in 5 or 4 dimensions with $N=2$. Vector multiplets in 6
dimensions do not contain scalars. When reducing to 3 dimensions, the
vectors become dual to scalars (we can perform duality transformations as
we are just considering  kinetic terms here, and we can thus restrict to
Abelian vectors). Therefore the multiplet in 3 dimensions is dual to a
multiplet with only scalars: the hypermultiplet. For hypermultiplets, the
spacetime dimension is not really relevant as there are no vectors, and
thus the results for hypermultiplets are the same for any dimension. In
the picture it is convenient to consider them in 3 dimensions because of
the dimensional reduction that we just described. With real scalars in the
vector multiplets in 5 dimensions, these geometries are real geometries,
those in 4 dimensions are K\"{a}hlerian, and the hypermultiplets lead to 3
complex structures. Furthermore, we can distinguish between theories that
appear in rigid supersymmetry, and those in supergravity. This leads to
the overview in the upper part of table~\ref{vpro-tbl:SpecGeom}.
\begin{table}
\begin{tabular}{|c|ccc|}
\hline
   & $D=5$ vector multiplets & $D=4$ vector multiplets & hypermultiplets \\
\hline
 rigid & affine  & affine &  \\
 (affine) & very special real & special K\"{a}hler & hyperk\"{a}hler  \\ \hline
 local & (projective) & (projective) &   \\
 (projective) & very special real & special K\"{a}hler  & quaternionic-K\"{a}hler \\
\hline \multicolumn{4}{r}{
\epsfxsize=13cm
 \epsfbox{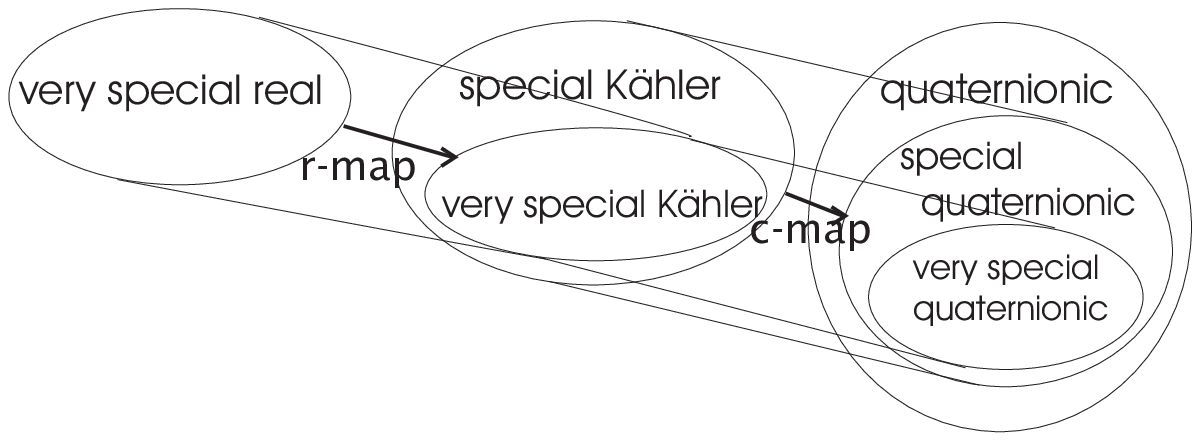} 
 }
\end{tabular}
  \caption{\it Geometries from supersymmetric theories with 8 real
  supercharges, and the connections provided by the ${\bf r}$-map and the
  ${\bf c}$-map.   }\label{vpro-tbl:SpecGeom}
\end{table}
The geometries that are related to rigid supersymmetry have been called
`affine' in the mathematics
literature~\cite{vpro-Freed:1997dp,vpro-Alekseevsky:1999ts}, while those
for supergravity are called `projective' (and these are the default, in
the sense that e.g. special K\"{a}hler refers to the geometry that is found
in supergravity. The analogous manifolds with 3 complex structures got
already a name in the literature.

The name `projective' versus `affine' can be understood from the
construction of the manifolds in supergravity using superconformal tensor
calculus. We saw already (see table~\ref{vpro-tbl:fieldsConf}) how the
real very special manifolds are obtained starting from $(n+1)$ vector
multiplets. Before any gauge fixing, these are just real manifolds with a
dilatational invariance. This manifold has therefore a cone structure,
with $C_{IJK}h^Ih^Jh^K$ as the radial coordinate. The physical scalars of
the supergravity theory are thus defined modulo this dilatational
scaling. The manifolds that occurs in supergravity can thus be seen as a
projective space of dimension $n$.

Similarly, to construct special K\"{a}hler geometry, one starts in 4
dimensions with the couplings as they occur in rigid supersymmetry,
demanding the presence of a superconformal symmetry. Again, the manifold
has a cone structure, and the dilatational gauge condition selects a
submanifold at fixed radius. In this case, the superconformal group
contains a $U(1)$ invariance and the manifold at fixed radius is a
`Sasakian manifold' of dimension $2n+1$, if this $U(1)$ is not gauged. In
conformal supergravity the $U(1)$ is local and eliminates one more
scalar. The gauge field of this $U(1)$, which is an auxiliary field in
the superconformal tensor calculus (similar to $V_{\mu i}{}^j$ in
table~\ref{vpro-tbl:fieldsConf}), becomes by its field equation the $U(1)$
connection on the K\"{a}hler manifold. The final manifold in super-Poincar\'{e}
has then non-trivial $U(1)$ curvature (and will be a Hodge-K\"{a}hler
manifold).

The construction for quaternionic manifolds is similar, as has been
demonstrated recently in 4 dimensions in~\cite{vpro-deWit:2001dj}. One
starts then from hyperk\"{a}hler cones. The dilatational gauge choice leads
to a tri-Sasakian manifold of dimensions $4r+3$ for ungauged $SU(2)$. The
$SU(2)$ gauge fields of the Weyl multiplet get by their field equations
the value $V_{\mu i}{}^j=\partial _\mu q^X \omega _{Xi}{}^j$, using the
$SU(2)$ connection that we had in the previous section. The $SU(2)$
curvature is thus non-zero as required by (\ref{vpro-RlambdaJ}).

Dimensional reduction gives a mapping between these manifolds. These
mappings have been called the $\bf c$-map (from special K\"{a}hler to special
quaternionic)~\cite{vpro-Cecotti:1989qn}, and the $\bf r$-map (from very
special real to very special K\"{a}hler)~\cite{vpro-deWit:1992nm}. They are
represented in the lower part of table~\ref{vpro-tbl:SpecGeom}.
Dimensional reduction of a manifold in 5 dimensions gives a 4-dimensional
theory. But the 4-dimensional theories that can be obtained in this way,
are only a subset of all 4-dimensional theories. The table shows the
structure in the names given to various classes of manifolds. Very
special K\"{a}hler manifolds are a subset of all special K\"{a}hler manifolds.
The quaternionic manifolds that are in the image of the $\bf c$-map are
the special quaternionic manifolds, and those in the image of the ${\bf
c}{\scriptstyle\circ}{\bf r}$-map are the very special quaternionic
manifolds. It is remarkable that nearly all the homogeneous quaternionic
manifolds are very special quaternionic manifolds~\cite{vpro-deWit:1992nm}
(The only non-special homogeneous quaternionic manifolds are the
quaternionic projective spaces).

\subsection{Gauging of isometries and the consequences}

Having vectors in the vector multiplets (and one graviphoton), these can
be used to gauge extra symmetries. The `extra' refers here to the fact
that these do not belong to the super-Poincar\'{e} or the superconformal
algebra. However, at the end, in supergravity, we do not have a strict
mathematical algebra of symmetries, but a soft algebra. This means that
there are structure functions rather than structure constants. These
functions depend on the fields in the theory. In this way the `extra'
gauge group can appear in the anticommutator of two supersymmetries with
structure functions depending on the fields. When these have a non-zero
expectation value, central charges appear. The generators of these gauge
group may act also on the fields of the quaternionic manifold.

In supersymmetry, such a gauging has three consequences
(\cite{vpro-Fre:2001jd} gives general properties of gauging in
supergravity). The first is that new terms appear in the supersymmetry
transformations of the fermions (proportional to a gauge coupling
constant). Secondly, the scalar potential is completely determined by the
gauging (this is true for theories with 8 supercharges and more).
Finally, gauged $R$-symmetry leads to a cosmological constant.
$R$-symmetry is the symmetry that rotates the supersymmetries and was
mentioned in table~\ref{vpro-tbl:sconfalg}. In $N=2$, $D=5$ it is
$SU(2)$. We obtain gauged $R$-symmetry if the extra gauge group contains
an $SU(2)$ or a $U(1)$ subgroup thereof. This extra gauge group mixes
with the $R$-symmetry. In the superconformal approach this is due to the
gauge fixing of the $SU(2)$ of the superconformal algebra by fixing the
scalars in the compensating hypermultiplet, see
table~\ref{vpro-tbl:fieldsConf}. The latter in general also transform
under the gauge symmetry (see~\cite{vpro-Wit:2001bk} for the structure of
the symmetries in the hyperk\"{a}hler cone). The remaining gauge group is a
diagonal subgroup of the superconformal $SU(2)$ and the extra gauge
group. The gauge fields of the extra gauge group then gauge the symmetry
that acts the supersymmetries (and on the gravitini). In that case we use
the terminology `gauged supergravity'. We will do this explicitly below
for a $U(1)\subset SU(2)$. The formulae can also be used for gauging the
full $SU(2)$ as has been done in~\cite{vpro-Gunaydin:2000ph}. Gauged
$R$-symmetry induces a cosmological constant\footnote{To have really a
`constant', one still has to assume that there is a solution such that the
scalars that determine the value of the potential are constant.}, which
is proportional to the square of the gauge coupling constant (the group
theoretical argument was repeated in~\cite{vpro-Bergshoeff:2000ii}).

The vectors are in the adjoint of the gauge group, and supersymmetry then
implies the same for the scalars $h^I$ in the $(n+1)$ dimensional space.
If the constraint (\ref{vpro-Chhh1}) is compatible with this, i.e.
$C_{L(IJ}f^L_{K)M}=0$, then the group can be gauged. It is for these
non-Abelian theories that tensor multiplets give extra
possibilities~\cite{vpro-GunZag,vpro-AnnaGianguido}.

In order that these symmetries can act on the hypermultiplet, the
quaternionic manifold should have isometries. Thus, we suppose that there
are Killing vectors $K^X_\alpha(q) $ that determine transformations of the
scalars $q^X$, and $\alpha $ denotes the different isometries. In
general, only a subset of these can be gauged. We need for each one a
gauge vector. Therefore, the appropriate index is $I$, labeling the
vectors (see table~\ref{vpro-tbl:superPoinc}), and the gauged isometries
are determined by $K_I^X(q)$. In quaternionic geometry, the isometries are
determined by a triplet of prepotentials $\Blue{P_I^r(q)}$:
\begin{equation}
  \Blue{{\cal R}_{XY}^r K_I^Y= D_X P_I^r}\,,\qquad D_X \Blue{P_I^ r}\equiv
\partial _X\Blue{P_I^ r}+2\varepsilon ^{rst}\omega _X^s\Blue{P_I^t} \,.
 \label{vpro-constrPrepot}
\end{equation}
These can be solved as well for the Killing vectors, or for the
prepotentials:
\begin{equation}
  \Blue{K^Z_I=-\ft43 {\cal R}^{r\,ZX}  D_X P_I^r\,,\qquad
  P_I^r=\frac1{2r} {\cal R}^{r\,XY}  D_X K_{IY}}\,.
 \label{vpro-solConstrP}
\end{equation}
The latter equation~\cite{vpro-ACCVP} is obviously only true for $r\neq
0$. If there are no physical hypermultiplets ($r=0$), then $P_I^r$ are
just some constants. In the superconformal approach they determine the
action of the symmetry on the compensating hypermultiplet. These are the
analogues of the Fayet--Iliopoulos terms. For $r>0$ there is thus no
Fayet--Iliopoulos term possible~\cite{vpro-Ceresole:2001wi}. In rigid
supersymmetry, the $SU(2)$ curvatures vanish, and $P_I^r$ are again
arbitrary constants or `Fayet--Iliopoulos terms'.

The above quantities determine the modified supersymmetries. For that
purpose one defines `dressed' Killing vectors and `dressed' prepotentials:
\begin{equation}
  P^r \equiv  \Red{h^I(\phi)} \Blue{P_{I}^{r}(q)}\,,\qquad
K^X\equiv \Red{h^I(\phi)} \Blue{K_{I}^X(q)}\,.
 \label{vpro-dressedKP}
\end{equation}
These thus depend as well on the scalars of the vector multiplets as on
those of the hypermultiplets, but in a well-structured way. The
supersymmetry transformations of the fermions are
then~\cite{vpro-AnnaGianguido} (bosonic terms only)
\begin{eqnarray}
\delta _\epsilon \psi _{\mu i}  & = & { D}_\mu (\omega )\epsilon_i+
\ft{1}{4\sqrt{6}}\rmi \left( \gamma_{\mu \nu \rho } - 4 g_{\mu \nu }
\gamma_\rho  \right)\epsilon_i h_I F^{\nu \rho \,I}
-\ft{1}{\sqrt{6}}\rmi g\gamma _\mu P_i{}^j\epsilon _j\,, \nonumber\\
 \delta _\epsilon \Red{\lambda _i^x}&=&-\ft{1}{2}\rmi(\not\!\!{D} \Red{\phi
^x})\epsilon _i +\ft{1}{4}h_I^{x}\gamma ^{\,\mu \nu }\epsilon _iF_{\mu \nu
}^I
-g\sqrt{\ft32} \epsilon ^j\partial _xP_i{}^j\,,\nonumber\\
 \delta_\epsilon \Blue{\zeta^A} &=&f^{Ai}_X\left[  \ft{1}{2}\rmi (\not\!\!{D}
  \Blue{q^{X}})\epsilon _i - g\ft14\sqrt{6}\epsilon _iK^X\right]\,.
  \label{vpro-fermiontransf}
\end{eqnarray}
We used here notations of very special geometry:
\begin{equation}
  h_I\equiv C_{IJK}h^Jh^K\,,\qquad h_I^x=\sqrt{\ft32}g^{xy}\partial _y
  h^I\,, \qquad   h^I_x\equiv -\sqrt{\ft32}\partial _xh^I\,.
 \label{vpro-hIxnot}
\end{equation}
Notice that the gauging produced an extra scalar-dependent term for each
of the fermions, apart from covariantizations depending on the gauge
vectors:
\begin{equation}
 D_\mu \epsilon_i= \ldots -gA_\mu ^I P_{Ii}{}^j\epsilon _j\,,\quad D_\mu \phi
 ^x=
 \ldots +gA_\mu ^I\sqrt{\ft32}h_Kf^K_{JI}h^{Jx}\,,\quad
 D_\mu q^X=
 \ldots +gA_\mu ^IK_I^X\,.
 \label{vpro-Dmu}
\end{equation}
Finally, as usual in supersymmetry, the scalar potential is a square of
the transformation laws of the fermions. We have here:
\begin{equation}
  V=g^2\left[ - 4 P^r P^r +3 (\partial _xP^r)(\partial ^x P^r)
 +\ft34K^XK_X\right].
 \label{vpro-Vscalar}
\end{equation}
A general form for scalar potentials has been put forward, guaranteeing
stability~\cite{vpro-Boucher:1984yx,vpro-Townsend:1984iu,vpro-Skenderis:1999mm}:
\begin{equation}
  V=g^2\left(- 6 \ForestGreen{W }^2+\ft92 g^{\Lambda\Sigma}\partial_\Lambda
 \ForestGreen{W }\partial_\Sigma \ForestGreen{W }\right) ,
 \label{vpro-Vstable}
\end{equation}
where $\Lambda ,\Sigma $ run over all the scalars, and $W$ is some
`superpotential'. To make the transition from (\ref{vpro-Vscalar}), we
first split the dressed prepotential in a norm $W$, which is identified as
superpotential, and a phase $Q^r$:
\begin{equation}
  P^r=\sqrt{\ft32} \ForestGreen{W }\Blue{Q^r}\,,\qquad \Blue{Q^rQ^r}=1\,.
 \label{vpro-Pnormphase}
\end{equation}
The phase determines which $U(1)$ subgroup of the $SU(2)$ $R$-symmetry is
gauged. Then it turns out~\cite{vpro-Ceresole:2001wi} that we can write
the potential as (\ref{vpro-Vstable}) if the derivative of the phase with
respect to the scalars of the vector multiplets is zero:
\begin{equation}
  \Red{\partial _x} \Blue{Q^r}=0\,.
 \label{vpro-dxQr0}
\end{equation}
An equivalent condition was found in~\cite{vpro-DW}. This condition is
automatic if there are no hypermultiplets (the prepotentials are then
constants) or if there are no vector multiplets. However, if one has
vector- and hypermultiplets then this is in general only satisfied on a
submanifold of the total scalar manifold. We will see below that this
condition is required also for BPS solutions of the theory.

\section{Smooth solutions: RS, flows and attractors}

\subsection{BPS conditions}
We now look for bosonic solutions with vanishing vectors, a metric of the
form (\ref{vpro-metric}), and preserving some amount of
supersymmetry~\cite{vpro-Ceresole:2001wi}. As the fermions are zero in
such solutions, their transformation laws should vanish too. Therefore we
investigate the vanishing of (\ref{vpro-fermiontransf}) in this
background. The transformation of $\psi _5$ determines the
$x^5$-dependence of the Killing spinors $\epsilon ^i(x)$, and we can
neglect this further. In the transformation of the $\psi _{\underline{\mu
}}$ appears a contribution of the spin connection. It gives the same
equation for each $\underline{\mu }=0,1,2,3$. This equation is still a
triplet of equations that can be split in the norm and the phase. These
equations are respectively (we use a prime to denote derivatives w.r.t.
$x^5$)
\begin{equation}
  g\ForestGreen{W}=\pm\ForestGreen{\frac{a'}{a}}\,,\qquad
 \rmi\gamma _5\epsilon _i=\mp \Blue{Q_i{}^j}\epsilon_j\,.
 \label{vpro-psimueqns}
\end{equation}
As we defined $W$ to be positive (a norm of a 3-vector), the sign in the
first equation depends on the sign of $a'/a$. The other signs then follow
from this one. The second equation is a projection on the preserved
supersymmetries. It implies that one half of the supersymmetries survives
(4 real supercharges, i.e. $N=1$ in 4 dimensions).

Note that we have considered only solutions that do not explicitly depend
on the coordinates $x^{\underline{\mu }}$. On `critical points' (see
below), other solutions are possible, doubling the number of preserved
supersymmetries (related to the $S$-supersymmetries in the dual conformal
theory).

The transformations of the gauginos can also be split in their norm and
phase as $SU(2)$ triplets. The phase gives again (\ref{vpro-dxQr0}), as we
announced already~\cite{vpro-Ceresole:2001wi}. The equation of the norm
and the equation for the hyperinos give a similar condition for all
scalars $\phi ^\Lambda $~\cite{vpro-DW}:
\begin{equation}
  \phi ^{\Lambda\, \prime}=\mp 3g\,g^{\Lambda \Sigma }
  \partial _\Sigma  \ForestGreen{W}\,.
 \label{vpro-BPSscalars}
\end{equation}

\subsection{Terminology of Renormalization group flow}

In the duality between the 5-dimensional theories and 4-dimensional
conformal theories, the scalars are dual to coupling constants. The
dependence on $x^5$ is denoted as a `flow'. The value of the warp factor
$a$ is dual to the energy scale. Therefore, $\beta$-functions, i.e.
logarithmic derivatives of the coupling constants to the energy, are
\begin{equation}
  \beta ^\Lambda =a\frac{\partial}{\partial a}\phi ^\Lambda =\frac{a }{
a'}\phi^{\Lambda \prime}=-3\frac{\partial ^\Lambda
\ForestGreen{W}}{\ForestGreen{W}}\,,
 \label{vpro-betafunct}
\end{equation}
where we used (\ref{vpro-psimueqns}) and (\ref{vpro-BPSscalars}). Critical
points ($\beta =0$) correspond thus to extrema of the superpotential and
at these points the scalars are constant. As well for a suitable RS
scenario as for a renormalization group flow, the end points of the flow
$x^5=\pm \infty $ should be such critical points.

Whether a critical point is a UV or an IR critical point depends on
whether the $\beta$-function decreases or increases while passing through
its zero. In the first case, the value of the scalars is attracted to
this point in the high-energy (large $a$) regime. In the second case they
are attracted to this point in the low-energy (small $a$) regime. The
type of critical point is thus determined by the matrix
\begin{equation}
  U_\Sigma{}^ \Lambda \equiv - \left.
\frac{\partial \beta ^\Lambda  }{\partial \phi ^\Sigma}\right|_{\beta =0}=
3\left. \frac{\partial _\Sigma \partial  ^\Lambda \ForestGreen{W}
}{\ForestGreen{W}}\right|_{\partial \ForestGreen{W} =0}\,.
 \label{vpro-defU}
\end{equation}
Positive eigenvalues imply that the point is a UV attractor for flows in
the direction of the corresponding eigenvector, while negative
eigenvalues indicate that it can be an IR attractor. The eigenvalues $u$
are the conformal dimensions in the dual theory and the scalar mass is
$M^2=u(u-4)\leq -4$, satisfying the Breitenlohner--Freedman
bound~\cite{vpro-Breitenlohner:1982bm}.

One can prove~\cite{vpro-ACCVP} a general formula for $U$:
\begin{equation}
  {\cal U}=\pmatrix{2\delta _x{}^y& -\frac1{W^2} (\partial _xK^Z){\cal J}_Z{}^Y\cr
  \frac1{W^2} {\cal J}_{XZ}\partial ^yK^Z
  &\frac32\delta _X{}^Y- \frac{1}{W^2}{\cal J}_X{}^Z{\cal L}_Z{}^Y}\,,
 \label{vpro-calUres}
\end{equation}
where the first entries are for the vector multiplets and the second for
the hypermultiplets, and ${\cal J}$ and ${\cal L}$ select respectively the
$SU(2)$ and $USp(2r)$ part of the dressed gauged isometry, defined by
\begin{equation}
 {\cal J}_{XY}\equiv 2 P^r{\cal R}^r_{XY} \,,\qquad  D_X K_Y={\cal J}_{XY}+{\cal
 L}_{XY}\,.
 \label{vpro-defcalJL}
\end{equation}
Note that with only vector multiplets, we have only the upper-left entry
of (\ref{vpro-calUres}), and thus only UV
attractors~\cite{vpro-KL,vpro-BC}. The trace of ${\cal J}{\cal L}$ is
zero, and thus the trace of the full matrix $U$ is $2n+6r>0$. Therefore
any critical point is UV in some directions.

The `attractor equations', which determine the conditions for critical
points, can be written as algebraic equations~\cite{vpro-Ceresole:2001wi}
\begin{equation}
  K^X  \equiv  \Red{h^I} \Blue{K_I^X}=0\,,\qquad
\Blue{P_I^r} =\Red{h_I}\,\Red{h^J}\Blue{P_J^r}\,.
 \label{vpro-attractoreqns}
\end{equation}
They have a group-theoretical meaning. The first one says that the
'dressed symmetry' at the critical point should be in the stability
subgroup of the isometry group. The second has $n$ components, as it is
trivial when multiplied with $h^I$. It says that at the critical point
the other $n$ symmetries should have the same $SU(2)$ content as the
dressed symmetry.

Also other equations can be understood in a group-theoretical way. The
value of the cosmological constant at these points is $-6W^2=-4P^rP^r$,
i.e. determined by the part of the gauging in the $SU(2)$ direction. On
the other hand, for an IR critical point, one needs that either the
lower-right entry of (\ref{vpro-calUres}) has to become negative, which
means that ${\cal L}$ has to be large, corresponding to a gauging in the
$USp(2r)$ part, or non-diagonal entries should be non-zero, which means
that Killing vectors should have parts that are not in the isotropy group
of that point.

These results allow to investigate the structure of flows for arbitrary
vector- and hypermultiplets in $N=2$, $D=5$ supergravity.
In~\cite{vpro-Ceresole:2001wi} couplings of 1 vector and 1 hypermultiplet
were considered, and a flow was found that generalizes the IR to UV flow
of~\cite{vpro-FGPW} with two arbitrary parameters. More general models
can be investigated easily due to the algebraic nature of the attractor
equations and their geometric significance.

\medskip
\section*{Acknowledgments.}

\noindent It was a pleasure to enjoy the friendly atmosphere of the
Karpacz school. My understanding of several aspects of the problems
mentioned in this text has grown by discussions with participants in this
school, especially I. Bars, E. Bergshoeff, G. Gibbons, J. Lukierski, K.
Stelle. This text summarizes work done in several collaborations. I thank
D. Alekseevsky, E. Bergshoeff, A. Ceresole, V. Cort\'{e}s, S. Cucu, G.
Dall'Agata, M. Derix, C. Devchand, B. de Wit, T. de Wit, R. Halbersma and
R. Kallosh for the fruitful collaborations. I also thank also J.
Gheerardyn and S. Vandoren for discussions.


\end{document}